\documentclass[conference]{IEEEtran}
\IEEEoverridecommandlockouts
\usepackage{cite}
\usepackage{amsmath,amssymb,amsfonts}
\usepackage{algorithmic}
\usepackage{graphicx}
\usepackage{textcomp}
\usepackage{xcolor}
\usepackage{multirow}
\usepackage{makecell}

\def\BibTeX{{\rm B\kern-.05em{\sc i\kern-.025em b}\kern-.08em
    T\kern-.1667em\lower.7ex\hbox{E}\kern-.125emX}}
\begin{document}

\title{Comprehensive Evaluation of Emergency Shelters in Wuhan City Based on GIS
}

\author{\IEEEauthorblockN{Tingyu Luo\textsuperscript{1}, Boheng Li\textsuperscript{2}, Jiahao Zhou\textsuperscript{3}, Qingxiang Meng\textsuperscript{3*}}
\IEEEauthorblockA{{\textsuperscript{1} School of Resource and Environmental Sciences, Wuhan University, Wuhan, China} \\
{\textsuperscript{2} School of Cyber Science and Engineering, Wuhan University, Wuhan, China} \\
\textsuperscript{3} School of Remote Sensing and Information Engineering, Wuhan University, Wuhan, China\\}
\textsuperscript{*} Corresponding author, e-mail: mqx@whu.edu.cn \\
}

\maketitle

\begin{abstract}
Emergency shelters, which reflect the city's ability to respond to and deal with major public emergencies to a certain extent, are essential to a modern urban emergency management system. This paper is based on spatial analysis methods, using Analytic Hierarchy Process to analyze the suitability of the 28 emergency shelters in Wuhan City. The Technique for Order Preference by Similarity to an Ideal Solution is further used to evaluate the accommodation capacity of emergency shelters in central urban areas, which provides a reference for the optimization of existing shelters and the site selection of new shelters, and provides a basis for improving the service capacity of shelters. The results show that the overall situation of emergency shelters in Wuhan is good, with 96\% of the places reaching the medium level or above, but the suitability level needs to be further improved, especially the effectiveness and accessibility. Among the seven central urban areas in Wuhan, Hongshan District has the strongest accommodation capacity while Jianghan District has the weakest, with noticeable differences.
\end{abstract}


\section{Introduction}
With the rapid development of cities and the continuous improvement of city governance systems, urban emergency management has become the focus of work. Emergency shelter is significant as it is a major component of city emergency resources and the important guarantee of people’s life safety. Nowadays, emergency shelters are required to be distributed evenly, safe and reliable, adaptive to local conditions, useful both in normal times and in an emergency as well as planned in the long term. However, at present, the construction of large emergency shelters has the problems of insufficient scale, uneven distribution, and incomplete functions.

Many current evaluations of emergency shelters focus on site selection and optimization, planning and design, spatial pattern, and serviceability. For instance, Japanese scholars Tetsuya and Kayoko\cite{b1} use Geographic Information Systems(GIS) and apply statistical methods and public open data related to population and emergency shelters. The present research aims to conduct a suitability analysis for the emergency shelter allocation quantitatively. Gu et al.\cite{b2} evaluated the rationality of the spatial layout of emergency shelters in the central urban area of Heyuan City from four aspects: accessibility, service area ratio, per capita accessible refuge area ratio, and population allocation gap, but did not combine the four aspects and conduct an overall evaluation. Xiong et al.\cite{b3} used AHP to evaluate the disaster reduction capability of the earthquake emergency shelters in Chaoyang District, Beijing, but did not evaluate the service capacity of the emergency places. Yin et al.\cite{b4} used AHP to evaluate the site selection of emergency shelters in Tianjin, combined with GIS spatial analysis methods to evaluate the satisfaction of 14 shelters in the central urban area. However, they did not establish a complete satisfaction evaluation system and did not compare and analyze the service capacity of the emergency shelters in each district.

Based on GIS spatial analysis methods, in this paper, we use AHP and Technique for Order Preference by Similarity to an Ideal Solution(TOPSIS) methods to evaluate the site selection and accommodation capacity of emergency shelters. The factors considered are relatively complete, and the analysis of population data is also combined. We clearly and intuitively demonstrated the current construction's shortcomings, thus facilitating future planning and layout of emergency shelters.

\section{Study area and data sources}

\subsection{Overview of the study area}
In this work, we study the emergency shelters in Wuhan City, Hubei Province, an important scientific and educational base, industrial base, and comprehensive transportation hub. It is located at the intersection of the golden waterway of the Yangtze River and the main artery of the Beijing-Guangzhou Railway and is known as the Nine Provinces Passage. Wuhan is also in an advantageous position in the economic geography circle of China and is the core city of the Yangtze River Economic Belt. Wuhan has a total area of 8,569 square kilometers, with a permanent population of about 11.45 million in 2021. Up to now, Wuhan has jurisdiction over 13 districts, including Jiang'an District, Jianghan District, Qiaokou District, Hanyang District, Wuchang District, Qingshan District, Hongshan District, Caidian District, Jiangxia District, Huangpi District, Xinzhou District, Dongxihu District, and Hannan District. 

\subsection{Data sources}
The data of Wuhan's administrative divisions, traffic network, water system, etc., are from OpenStreetMap\footnote{https://www.openhistoricalmap.org} and were pretreated through ArcGIS software. The directory and relevant information of Wuhan emergency shelters, major hazard sources, hospitals, fire stations, and public security organs originate from Wuhan public data open platform\footnote{https://data.wuhan.gov.cn}. The coordinates of these places are obtained through the Baidu map coordinate picking system and converted to the earth coordinates. The DEM data of Wuhan comes from geospatial data cloud\footnote{https://www.gscloud.cn}. The data on Wuhan City area, resident population, population density, and other data comes from the Wuhan City Statistical Yearbook (2021)\footnote{http://tjj.wuhan.gov.cn/tjfw/tjnj/}.

\section{Suitability evaluation of emergency shelters}

\subsection{Evaluation system}
\begin{figure*}[htpb]
\centerline{\includegraphics[width=6in]{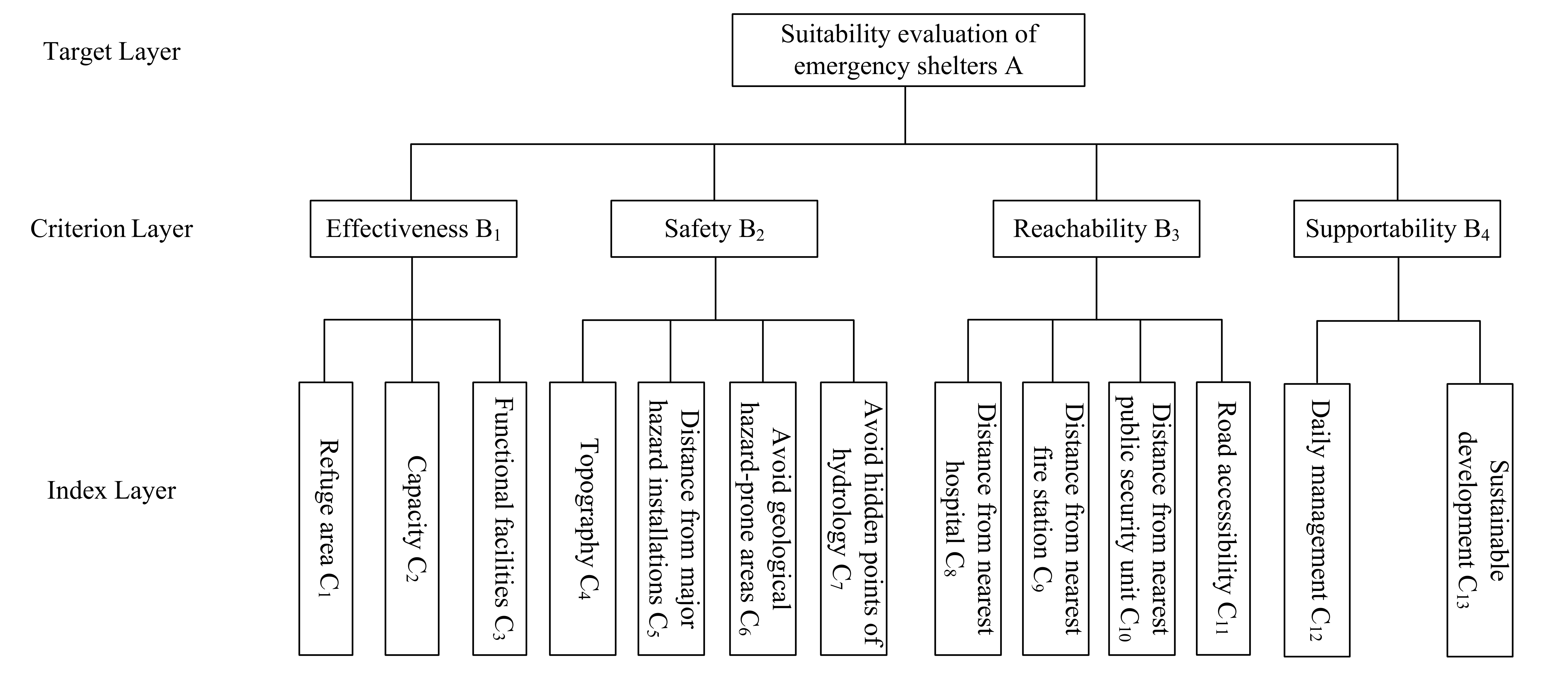}}
\caption{The AHP model for the suitability evaluation of emergency shelters}
\label{fig1}
\end{figure*}

Based on the functions and significance of emergency shelter, the suitability evaluation of emergency shelter is decomposed into target layer, criterion layer, and index layer by AHP. Combined with Construction Standard of Urban Community Emergency Shelters (JB 180-2017) and Wuhan Emergency Project Management Measures, the evaluation indexes are selected from the four dimensions of effectiveness, safety, reachability and supportability, and finally the analytic hierarchy process model for the suitability evaluation of emergency shelters in Wuhan is constructed (see Fig.~\ref{fig1}).

\subsection{Calculation of the model}
In this study, following the working methods of Chu et al.\cite{b5}, the relative importance of each element are obtained and then the judgment matrices of each criterion layer and index layer are constructed, and the weights are calculated after passing the consistency test (shown in Table 1).

\begin{table}[htbp]
\caption{Each evaluation index and its weight value}
\centering
\begin{tabular}{|c|c|}
\hline
Index Layer & Weight(W) \\ \hline
Refuge area C$_{\mbox{\scriptsize 1}}$ & 0.0930 \\ \hline
Capacity  C$_{\mbox{\scriptsize 2}}$  & 0.0930 \\ \hline
Functional facilities C$_{\mbox{\scriptsize 3}}$  & 0.0465 \\ \hline
Topography C$_{\mbox{\scriptsize 4}}$  & 0.0841 \\ \hline
Distance from major hazard installations C$_{\mbox{\scriptsize 5}}$  & 0.1780 \\ \hline
Avoid geological hazard-prone areas C$_{\mbox{\scriptsize 6}}$  & 0.0519 \\ \hline
Avoid hidden points of hydrology C$_{\mbox{\scriptsize 7}}$  & 0.0519 \\ \hline
Distance from nearest hospital C$_{\mbox{\scriptsize 8}}$  & 0.0525 \\ \hline
Distance from nearest fire station C$_{\mbox{\scriptsize 9}}$  & 0.0597 \\ \hline
Distance from nearest public security unit C$_{\mbox{\scriptsize {10}}}$  & 0.0294 \\ \hline
Road accessibility C$_{\mbox{\scriptsize {11}}}$  & 0.1362 \\ \hline
Daily management C$_{\mbox{\scriptsize {12}}}$  & 0.0928 \\ \hline
Sustainable development C$_{\mbox{\scriptsize {13}}}$  & 0.0310 \\ \hline
\end{tabular}
\end{table}

\subsection{Calculation of the comprehensive index}
Calculate the comprehensive indexes of emergency shelters in Wuhan, and get the results of suitability evaluation according to the comprehensive index. The calculation process mainly includes:
\begin{enumerate}
\item Establish a geographic information database of emergency shelters in Wuhan City using ArcGIS.
\item Give a score to each index. C$_{\mbox{\scriptsize 3}}$, C$_{\mbox{\scriptsize {11}}}$ and C$_{\mbox{\scriptsize {13}}}$ were artificially graded according to the data of each site, and were given 2, 5 and 8 points respectively. Other indexes are divided into 10 levels by using natural breakpoint method in ArcMap, and 1-10 points are given according to the levels to minimize the influence of subjective factors.
\item Use the following formula to calculate the comprehensive index of each emergency shelter. 
$$P=\sum\nolimits_{i=1}^{13}W_iF_i,i=1,2,...,13$$
\item Conduct suitability evaluation. Divided the calculated comprehensive indexes into five grades (Table 2). The suitability evaluation results of emergency shelters in Wuhan are shown in Table 3 and Figure 2.
\end{enumerate}

\begin{table}[htbp]
\caption{Classification of suitability grades}
\centering
\begin{tabular}{|m{2.3cm}<{\centering}|m{1.2cm}<{\centering}|m{4cm}<{\centering}|}
\hline
Composite Index(P) & Grade(G) & Grade Description \\
\hline
8.001-10.000 & A & Better suitability, it is reasonable in terms of effectiveness, safety, reachability and supportability. It is a good emergency shelter and needs to be maintained and improved.\\
\hline
6.001-8.000 & B & Good suitability, it needs to maintain good aspects, and make targeted improvements of deficiencies.\\
\hline
4.001-6.000 & C & Medium suitability, it basically meets the functional requirements of emergency shelter and needs further improvement.\\
\hline
2.001-4.000 & D & Poor suitability, it still has great  room for improvement.\\
\hline
0.000-2.000	& E	& Poorer, it is not suitable to be used as an emergency shelter.\\
\hline
\end{tabular}
\end{table}

\begin{table*}[htbp]
\caption{Score of each criterion layer and comprehensive index}
\centering
\begin{tabular}{|c|c|c|c|c|c|c|}
\hline
Administrative Division & Name & B$_{\mbox{\scriptsize 1}}$ & B$_{\mbox{\scriptsize 2}}$ & B$_{\mbox{\scriptsize 3}}$ & B$_{\mbox{\scriptsize 4}}$ & P \\
\hline

Jiang'an District & Baiduting Garden & 1.800 & 8.088 & 7.161 & 5.000 & 5.986 \\
\hline

\multirow{3}*{JiangHan District} & Zhongshan Park & 4.800 & 4.459 & 8.305 & 8.000 & 6.045\\
\cline{2-7}
~& Changqing Park & 6.000 & 6.345 & 6.753 & 8.000 & 6.583 \\
\cline{2-7}
~& International Convention and Exhibition Center &2.600 & 3.858 & 9.811 & 8.000 & 5.732 \\
\hline

\multirow{3}*{Qiaokou District} & Qiaokou Park & 4.400 & 6.467 & 8.026 & 7.250 & 6.516\\
\cline{2-7}
~& Zhuyehai Park & 4.800 & 3.345 & 4.667 & 2.750 & 3.977 \\
\cline{2-7}
~& Hubei University of Plice & 2.600 & 6.831 & 4.519 & 7.250 & 5.257 \\
\hline

\multirow{2}*{Hanyang District} & Qintai Square & 4.800 & 4.230	& 5.449	& 8.000	& 5.168\\
\cline{2-7}
~& Sports Training Base	& 4.200	& 6.440	& 7.148 & 5.750	& 6.030 \\
\hline

\multirow{3}*{Wuchang District} & Shahu Park & 5.800 & 5.716 & 4.459 & 8.000 & 5.669\\
\cline{2-7}
~& Integrity Park & 4.600 & 3.973 & 6.141 & 8.000 & 5.220 \\
\cline{2-7}
~& Wuhan Conservatory of Music & 5.200 & 4.804 & 7.106 & 5.000 & 5.560\\
\hline

\multirow{2}*{Qingshan District} & Qingshan Park & 8.600 & 7.433 & 5.189 & 8.000 & 7.151\\
\cline{2-7}
~& Heping Park & 8.600 & 7.548 & 4.180 & 8.000 & 6.913 \\
\hline

Hongshan District & Emergency shelter in Hongshan Square & 5.600 & 5.115 & 5.426 & 5.750 & 5.393 \\
\hline

\multirow{2}*{Dongxihu District} & Wuhan Dongxihu Vocational Technical School & 3.600 & 7.601 &3.478 & 5.000 & 5.204\\
\cline{2-7}
~& Wuhuan Square & 4.600 & 6.433 & 4.010 & 8.000 & 5.528 \\
\hline

\multirow{3}*{Hannan District} & Zhujiashan Park & 4.600 & 6.061 & 2.060 & 8.000 & 4.850\\
\cline{2-7}
~& Shamo Riverbank Park	 & 9.200 & 6.183 & 1.321 & 8.000 & 5.759 \\
\cline{2-7}
~& Hubei Land Resources Vocational College & 5.800 & 5.655 & 3.131 & 5.000 & 4.907\\
\hline

\multirow{2}*{Caidian District} & Emergency shelter in Wenti Square	& 4.400	& 5.222	& 3.449	& 8.000	& 4.883\\
\cline{2-7}
~& Emergency shelter in Riverbank Park & 7.600	& 7.582	& 5.817	& 8.000	& 7.147 \\
\hline

\multirow{3}*{Jiangxia District} & Civic Leisure Center	& 6.000	& 6.372	& 2.808	& 7.250	& 5.404\\
\cline{2-7}
~& Xiong Tingbi Park & 7.800 & 7.885 & 2.298 & 8.000 & 6.327 \\
\cline{2-7}
~& Century Square & 5.800 & 7.088 & 3.635 & 8.000 & 5.942\\
\hline

\multirow{2}*{Huangpi District} & Erlongtan Park & 3.000 & 7.946 & 3.032 & 5.750 & 5.159\\
\cline{2-7}
~& Huangpi Square	& 4.200	& 5.939	& 5.689	& 8.000	& 5.720 \\
\hline

Xinzhou District &  Human Defense Evacuation Base & 4.000 & 9.345 & 4.807 & 7.250 & 6.582 \\
\hline

\end{tabular}
\end{table*}

\begin{figure}[htbp]
\centerline{\includegraphics[width=3in]{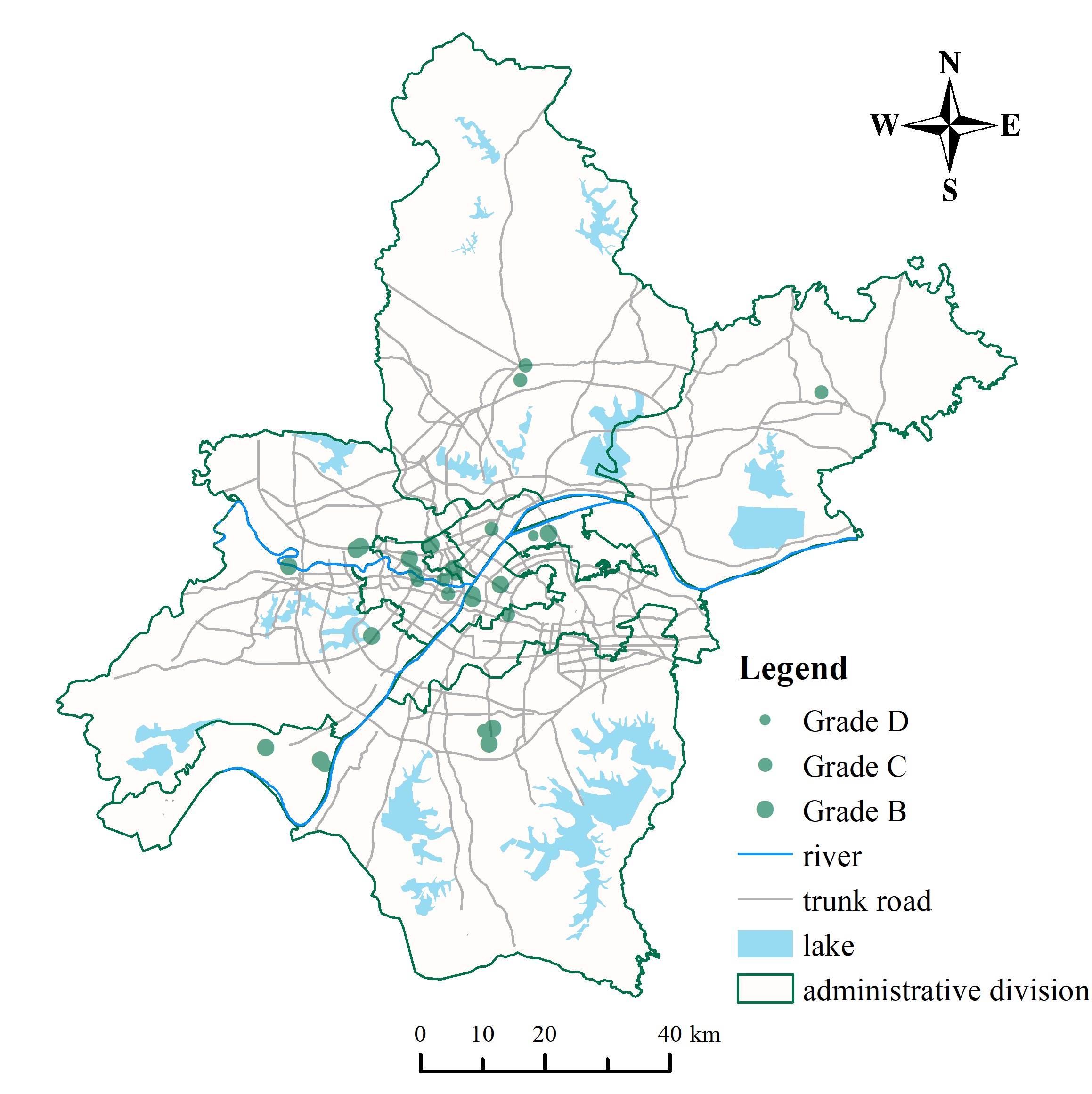}}
\caption{The suitability evaluation results of emergency shelters}
\label{fig}
\end{figure}

\subsection{Results analysis}
Overall, the emergency shelters in Wuhan are in good condition. Among the 28 selected emergency shelters, 14 are suitable, 13 are medium, and one is sub-standard. Qingshan Park, located in Qingshan District, has the highest suitability level, while the Zhuyehai Park in Qiaokou District is the lowest.

From the effectiveness perspective, more than 57 percent of emergency shelters scored less than 5 points, indicating that they have limited ability to provide help to surrounding citizens in response to emergencies, which is mainly limited by the number of people they can accommodate. Baibuting Garden, located in Jiang'an District, needs to be strengthened because of its small area, the small number of people, and low level of effectiveness.

In terms of safety, the overall situation is quite good. The emergency evacuation places of Lianzheng Park, International Convention and Exhibition Center, Zhuyehai Park, and Wenti Square have potential safety hazards because they are close to major hazard sources such as hazardous chemical enterprises, which means it is necessary to make a response plan for major hazard sources. Human Defense Evacuation Base is located in a flat and high terrain, which effectively avoids geological and hydrological hidden danger points and has a high level of safety.

Regarding reachability, emergency shelters located in the central urban area are generally better, mainly because of their strong road evacuation capacity, good road connectivity, and many surrounding hospitals, police stations, and fire stations. The emergency evacuation places of Hubei University of Plice, Zhujiashan Park, Shamao Riverbank Park and Wenti Square are relatively far away from the hospital, which is not conducive to rapid medical assistance; The stadium of the Hubei Land Resources Vocational College, Erlongtan Park, Huangpi square and Xinzhou District civil air defense evacuation base are far away from the fire station, which is not conducive to the development of rescue operations. For the above emergency shelters, small medical stations and firefighting strongholds can be added, with basic emergency rescue capacity.

From the perspective of supportability, the overall situation of emergency shelters in Wuhan is good, except for Zhuyehai Park, reaching 5 points or more. For the above places, regular maintenance shall be carried out, daily management shall be strengthened, and a complete emergency plan shall be established to realize the sustainable utilization of materials and materials used in shelters.

\section{Capacity evaluation of emergency shelters}

\subsection{Evaluation system}
Based on the service capacity requirements of urban emergency shelters, this study evaluates the capacity of emergency shelters in Wuhan from three dimensions of satisfaction, effectiveness, and applicability by using TOPSIS according to the Overall Emergency Plan for Emergencies of Wuhan(see Table 4).

\begin{table*}[htbp]
\caption{The evaluation index of capacity of emergency shelter}
\centering
\begin{tabular}{|m{1.3cm}<{\centering}|m{1.3cm}<{\centering}|m{5cm}<{\centering}|m{7cm}<{\centering}|m{1.2cm}<{\centering}|}
\hline
First index & First weight & Secondary index & The calculation method of secondary index & Secondary weight\\
\hline

\multirow{2}*{Satisfiability} & {\multirow{2}{*}{0.35}}& Total refuge area & The total area of the shelters & 0.5\\
\cline{3-5}
~&~& Total refuge population & Total number of people accommodated in the shelters &0.5 \\
\hline

\multirow{2}*{Effectiveness} & {\multirow{2}{*}{0.35}}& Effective range of refuge & The total area of effective service coverage of the shelters & 0.5\\
\cline{3-5}
~&~& Effective range of refuge & The total area of effective service coverage of the shelters &0.5 \\
\hline

\multirow{2}*{Applicability} & {\multirow{2}{*}{0.3}}& Average refuge area of a person & Ratio of total refuge area to permanent population & 0.5\\
\cline{3-5}
~&~& Average effective refuge area of a person & Ratio of total refuge area to effective refuge population &0.5 \\
\hline

\end{tabular}
\end{table*}

\subsection{Build the model}
Statistics on the basic situation of emergency shelters, and calculate the total refuge area, total refuge population, and average refuge area of a person, etc., of each administrative division (Table 5).

\begin{table*}[htbp]
\caption{Statistics of emergency shelters of central urban area}
\centering
\begin{tabular}{|m{1.7cm}<{\centering}|m{2cm}<{\centering}|m{1.6cm}<{\centering}|m{2cm}<{\centering}|m{1.6cm}<{\centering}|m{1.5cm}<{\centering}|m{1.6cm}<{\centering}|m{2.5cm}<{\centering}|}
\hline
Administrative Division	& Area of administrative division /km10\textsuperscript{2} & Permanent population /(10\textsuperscript{4}·cap) & Population density /(cap/ha) & Number of emergency shelters /pcs & Total refuge area /ha & Total refuge population /(10\textsuperscript{4}·cap)	 & Average refuge area of a person /m\textsuperscript{2} \\
\hline
Jiang'an District & 80.28 & 96.53 & 120.24 & 60	& 59.61	& 28.545 & 0.6175 \\
\hline
Jianghan District & 28.29 & 64.79 & 229.03 & 14	& 55.4522 & 50.9871	& 0.8559 \\
\hline
Qiaokou District & 40.06 & 66.67 & 166.42 & 62 & 133.301 & 56.8943 & 1.9994\\
\hline
Hanyang District & 111.54 & 83.73 & 75.06 & 76 & 173.2681 & 37.139 & 2.0694\\
\hline
Wuchang District & 64.58 & 110.22 & 170.67 & 100 & 111.5981	& 89.9178 & 1.0125\\
\hline
Qingshan District & 57.12 & 43.18 & 75.60 & 22 & 103.1 & 68.02 & 2.3877 \\
\hline
Hongshan District & 573.28 & 255.43	& 44.56	& 85 & 259.41 & 128.14 & 1.0156 \\
\hline
Central urban area & 955.15	 & 720.55 & 75.44 & 419	& 895.7394 & 459.6432 & 1.2431 \\
\hline

\end{tabular}
\end{table*}

Determine effective service radius and generate the effective service range. Emergency shelters are places for emergency evacuation and refuge that can be reached within 5-10 minutes after the disaster, considering the collapse of buildings and road damage caused by disasters, the speed for people to move to emergency shelters is set at 4km/h. In this paper, 667m, a 10min walking distance, is taken as the service radius, and the buffers of emergency shelters are generated by using the analytical buffer tool of ArcGIS (Fig. 3).

\begin{figure}[htbp]
\centerline{\includegraphics[width=3.5in]{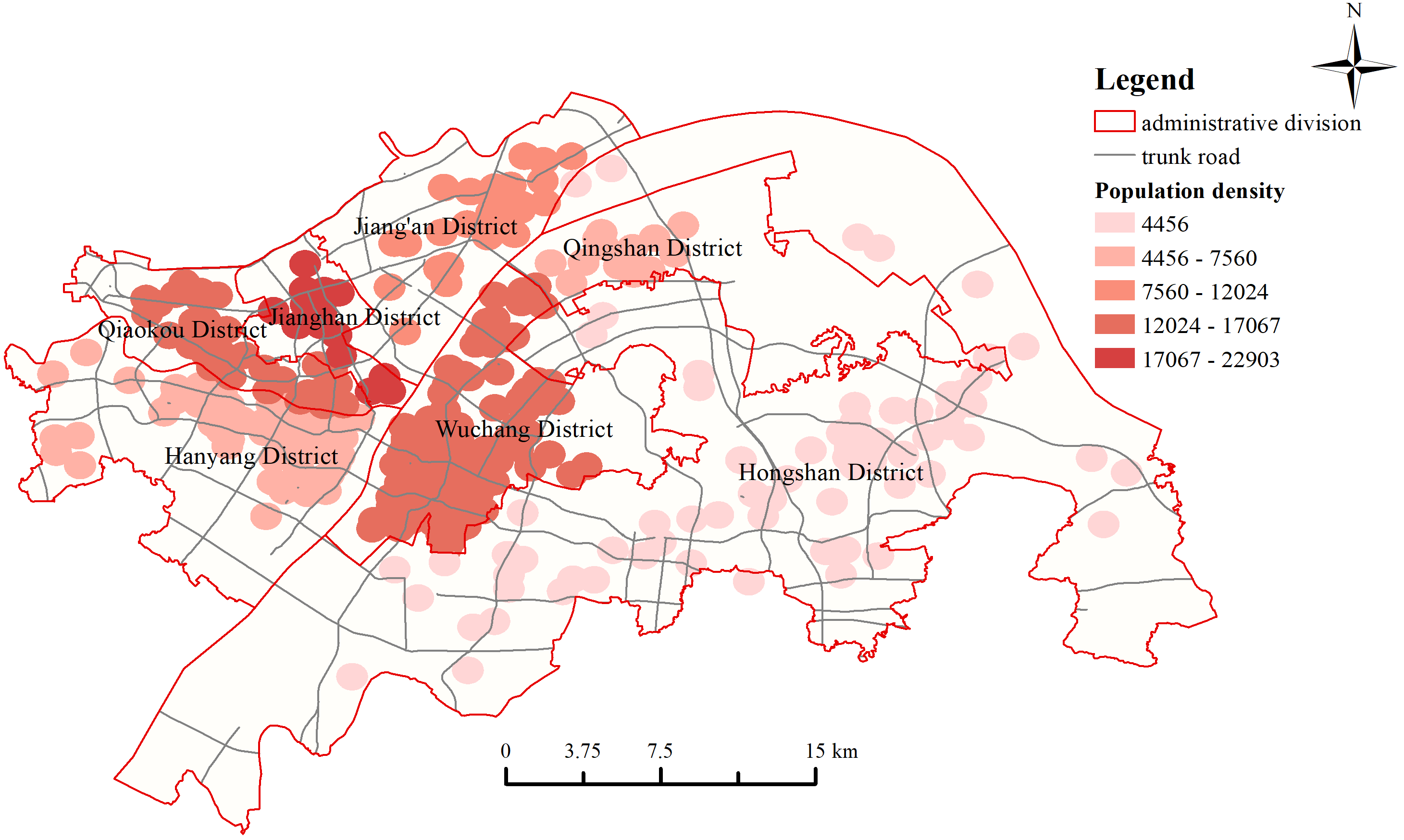}}
\caption{The suitability evaluation results of the central urban areas}
\label{fig}
\end{figure}

Determine the coverage area of effective service. The buffer generated above is used to calculate the area of the effective refuge range, and the population density distribution is superimposed with the buffer for geographical statistical analysis so as to calculate the effective refuge population of each district and further calculate the average effective refuge area of a person.

\subsection{Calculation of the model}
\begin{enumerate}
\item Use the following formula to standardize all secondary indexes and construct the standardization matrix of secondary indicators (Table 6).
$$z_{ij}=x_{ij}/\sqrt[]{\sum\nolimits_{i=1}^{7}x_{ij}^{2}} ,j=1,2,...,6$$
\begin{table*}[htbp]
\caption{Standardized results of evaluation indexes}
\centering
\begin{tabular}{|m{2.9cm}<{\centering}|m{2cm}<{\centering}|m{1.6cm}<{\centering}|m{2cm}<{\centering}|m{2cm}<{\centering}|m{1.64cm}<{\centering}|m{2.8cm}<{\centering}|}
\hline
Administrative Division	& Total refuge area & Total refuge population & Effective range of refuge	& Effective refuge population & Average refuge area of a person & Average effective refuge area of a person \\
\hline
Jiang'an District & 0.1567 & 0.1481	& 0.3401 & 0.3195 & 0.1491 & 0.0825 \\
\hline
Jianghan District & 0.1457 & 0.2645 & 0.0794 & 0.1420 & 0.2067 & 0.1727 \\
\hline
Qiaokou District & 0.3503 & 0.2951 & 0.3514	& 0.4570 & 0.4828 & 0.1290\\
\hline
Hanyang District & 0.4553 & 0.1927 & 0.4308	& 0.2527 & 0.4997 & 0.3032\\
\hline
Wuchang District & 0.2933 & 0.4664 & 0.5668	& 0.7559 & 0.2445 & 0.0653\\
\hline
Qingshan District & 0.2709 & 0.3528	& 0.1247 & 0.0737 & 0.5765 & 0.6189 \\
\hline
Hongshan District & 0.6817 & 0.6647	& 0.4818 & 0.1678 & 0.2452 & 0.6838 \\
\hline
\end{tabular}
\end{table*}
\item $Z^+=(0.6817,0.6647,0.5668,0.7559,0.5765,0.6838)$ is the optimal solution, and $Z^-=(0.1457,0.1481,0.0794,0.0737,0.1491,0.0653)$ is the worst according to the standardized matrix. 
\item Use the following formulas to calculate the weighted distance between each evaluation object and the optimal solution and the worst.
$$D_i^+=\sqrt[]{\sum\nolimits_{j=1}^{6}w_{j}(Z_j^+-z_{ij})^2}$$
$$D_i^-=\sqrt[]{\sum\nolimits_{j=1}^{6}w_{j}(Z_j^--z_{ij})^2}$$
\item Use the following formula to calculate the degree of closeness between the evaluation object and the optimal solution. The larger Si is, the higher the closeness degree is. The evaluation results are shown in Table 7 and Figure 4.
$$S_i=D_i^-/(D_i^++D_i^-)$$
\end{enumerate}

\begin{table}[htbp]
\caption{The evaluation results of the capacity}
\centering
\begin{tabular}{|m{2.6cm}<{\centering}|m{1cm}<{\centering}|m{1cm}<{\centering}|m{1cm}<{\centering}|m{1cm}<{\centering}|}
\hline
Administrative Division & D\textsuperscript{+} & D\textsuperscript{-}	& S	& Ranking \\
\hline
Hongshan District & 0.2797 & 0.4309	& 0.6063 & 1 \\
\hline
Wuchang District & 0.3274 & 0.3820 & 0.5385	& 2 \\
\hline
Hanyang District & 0.3437 & 0.2670 & 0.4372	& 3\\
\hline
Qiaokou District & 0.3382 & 0.2590 & 0.4337	& 4\\
\hline
Qingshan District & 0.4035 & 0.2895	& 0.4177 & 5\\
\hline
Jiang'an District & 0.4679 & 0.1501	& 0.2429 & 6 \\
\hline
Jianghan District & 0.4955 & 0.0736	& 0.1293 & 7 \\
\hline
\end{tabular}
\end{table}

\begin{figure}[htbp]
\centerline{\includegraphics[width=3.5in]{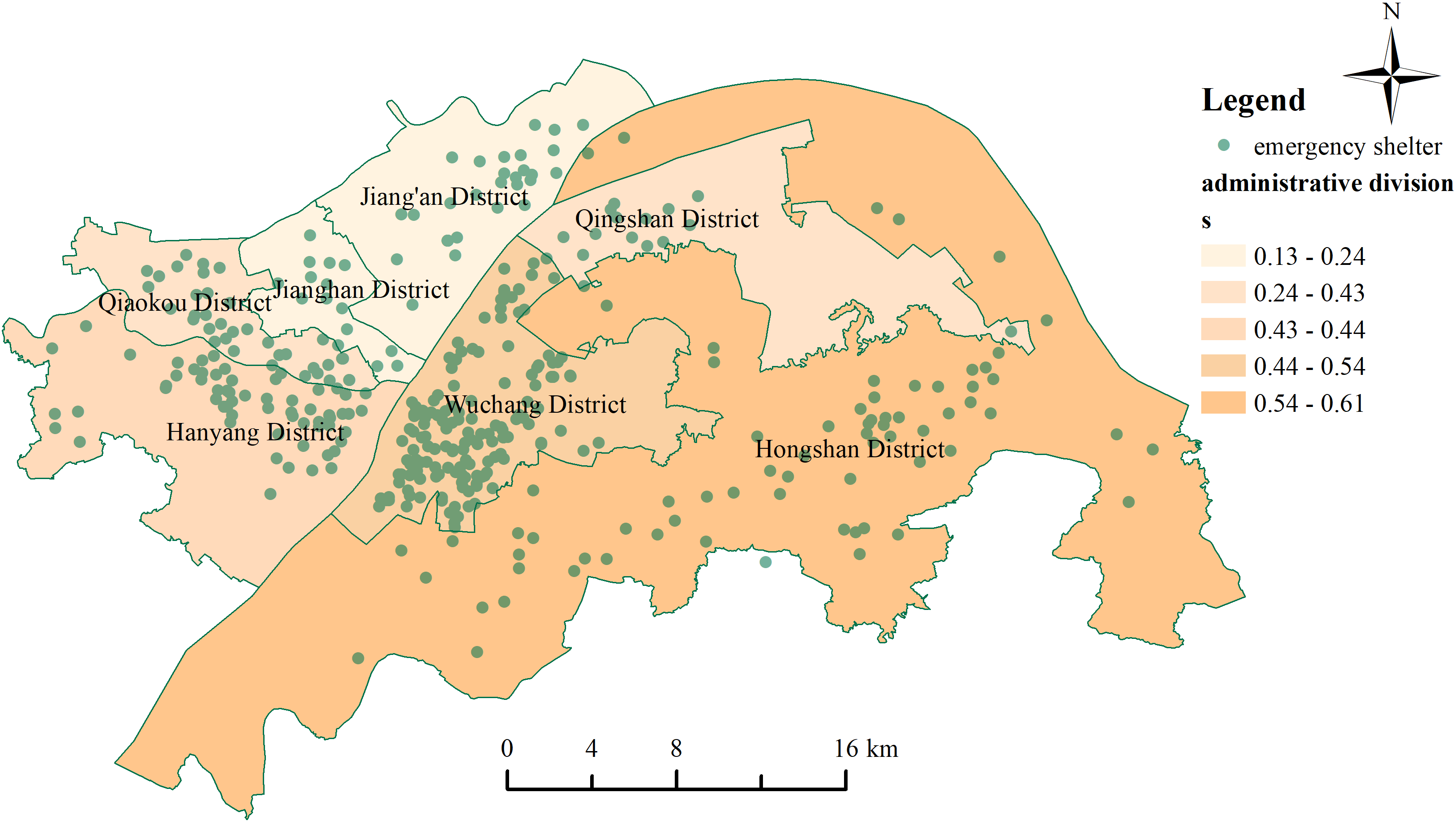}}
\caption{The results map of evaluation}
\label{fig}
\end{figure}

\subsection{Results analysis}
Combined with Figure3 and Figure 4, it can be seen that the distribution of emergency places in the central urban area of Wuhan is uneven. From Figure 4, it can be seen that Hongshan District has the strongest capacity of emergency shelters in the central urban area of Wuhan, and Jianghan District has the worst, and there is a noticeable gap. Combined with Table 6 and Table 7, it can be found that Jianghan District has a pronounced disadvantage in terms of effective refuge area, and Hongshan District has a significant advantage in three indicators: total refuge area, total refuge population, and average refuge area of a person, which can meet the basic needs of emergency evacuation. In addition, further calculations found that Qiaokou District and Wuchang District have a high proportion of effective service areas. In contrast, Hongshan district and Qingshan District have a high proportion of blind service areas, which need to strengthen the construction of emergency shelters. Hanyang District and Jiangan District have many emergency shelters. However, the proportion of effective service areas still has room for improvement, indicating that the spatial distribution of emergency shelters is not reasonable and needs to be optimized for site selection.

\section{Conlusion}

In this paper, we mainly discuss the construction of emergency shelters in Wuhan City. Based on principles and methods of spatial analysis of GIS, AHP and TOPSIS were used to conduct a separate and comprehensive evaluation of the suitability and capacity of the shelters to offer a reference to evaluating emergency shelters in other modern medium or large size cities. The refinement of the emergency management governance system and the modernization of governance capabilities are essential links in today's urban governance. The comprehensive capacity of emergency shelters reflects the emergency response capabilities of cities to a certain extent. Each city should scientifically organize, rationally distribute and establish multiple emergency shelters of different types and levels that are appropriate and safe according to its unique circumstances and possible disasters to satisfy its actual needs.

Results show that the emergency shelters in Wuhan need to be improved in the following three aspects. Firstly, it is necessary to improve the status of uneven distribution and an insufficient number of emergency shelters in non-central urban areas, make full use of urban land and space and balance the layout. Secondly, the governance and maintenance of the emergency shelters need to be improved. Parks, squares, stadiums, schools, and other public service facilities should be utilized according to local conditions. Those public infrastructures should be transformed or upgraded into emergency shelters in accordance with the requirements of upgrading by different levels and improving functions. Materials should be reserved for possible disasters, emergency escape instruction signs and emergency broadcasting facilities should be set up, and emergency plan management and emergency drill evaluation should be strengthened. Thirdly, micro-fire stations, medical stations, and emergency rescue teams should be set up in some emergency shelters, medical, fire-fighting and other resources should be rationally integrated and allocated. Unified management and deployment should be conducted as well.

\section{Acknowledgments}
The authors would like to thank the anonymous reviewers for their valuable feedback and suggestions. The authors would also like to express their gratitude to their colleagues and mentors who provided support and guidance throughout Tingyu's research and writing process. Boheng and Tingyu made equal technical contribution for this work.

\end{document}